\documentclass[aps,pra,showpacs,twocolumn,twocolumn]{revtex4}
\usepackage{graphicx}
\usepackage{color}
\usepackage{dcolumn}
\usepackage{bm}
\def\bra{\langle}
\def\ket{\rangle}
\def\<{\langle}
\def\>{\rangle}

\begin{document}
\title{Mathematical Theory of the Duality Computer in the Density Matrix Formalism}
\author{Gui Lu Long$^{1,2}$ }
\email{gllong@mail.tsinghua.edu.cn}
\address{$^1$ Key Laboratory for Quantum Information and Measurements and
Department of Physics, Tsinghua University, Beijing
100086, China\\
$^2$ Key Laboratory of Atomic and Molecular NanoSciences, Tsinghua University, Beijing
100084, China }
\date{\today }

\begin{abstract}
We give the mathematical theory of duality computer in the density matrix formalism.
This result complements the mathematical theory of duality computer of Gudder in the
pure state formalism.
\end{abstract}

 \maketitle

\section{Introduction}

Very recently, Gudder has given the mathematical theory for the duality computer
\cite{r1}. In his timely work, he has provided two descriptions, one is in the pure
state formalism and the other is in the mixed state formalism for the mathematics of
duality computer. The duality computer is a new type of computing machine based on the
general principle of quantum interference proposed in Ref.\cite{r2}. Notably, unsorted
database search problem may be solved by using only a single query, and all NP-complete
problems may have polynomial algorithms in the duality computer\cite{r2}.

In Ref.\cite{r1}, it has been pointed out that a paradoxical situation occurs with the
mixed state situation where the formalism gives a different result from the pure state
formalism, and in particular, the advantage of duality computer was lost in the mixed
state formalism.

In this paper, we give the mathematical theory of the duality computer in the density
matrix formalism. This density matrix formalism description is in accordance with
Gudder's pure state description, and hence solves the paradox.  Density matrix may have
different physical interpretations \cite{r3,r4}, and different interpretation may lead
to different results if issues concerning the fundamentals are involved. In this short
paper, we can either consider the density matrix arises from the average of many
computations using duality computers starting from different initial states, or
consider it as an improper mixed state arising from the average over other unseen
degrees of freedom. Here we just study the mathematical formalism and do not consider
fundamental issues where difference may arise.

\section{Divider and Combiner operators in Duality Computer}

Two important operations in the duality computer are the quantum wave divider and
combiner\cite{r2}. A quantum wave divider is an operation that divides a quantum wave
into many sub-waves of the same quantum system, and is represented graphically as
\begin{eqnarray}
\psi\longrightarrow\left\{\begin{array}{c}  p_1\psi\\
                      p_2\psi\\
                      \cdots\\
                      p_n\psi\end{array}\right.\label{divider}
                      \end{eqnarray}
where $p_i\ge 0$ and $\sum_{i=1}^n p_i=1$. The quantum wave combiner is the reverse
operation of quantum wave divider, and it is represented graphically as
\begin{eqnarray}
\left.\begin{array}{c}\psi_1\\
\psi_2\\
\cdots\\
\psi_n\end{array}\right\}\longrightarrow (\psi_1+\psi_2+\cdots+\psi_n)
\end{eqnarray}
Their mathematical descriptions have been given by Gudder as
\begin{eqnarray}
D_p\psi={1\over \|p\|} \oplus_{i=1}^n\left(p_i\psi\right), \end{eqnarray} for the
divider, and the quantum wave combiner is given by
\begin{eqnarray}
C_p(p_1\psi\oplus\cdots\oplus p_n\psi)=\|p\|\sum_{i=1}^n p_i\psi.
\end{eqnarray}
 where the subscript $p$ represents a probability distribution
$p=(p_1,\cdots,p_n)$ with $p_i\ge 0$ and $\sum_{i=1}^n p_i=1$. $\|p\|=\left(\sum
p_i^2\right)^{1/2}$. Then Gudder has shown that $D_p$ and $C_p$ are linear isometries
satisfying $C_p=D_p^{\dagger}$ where the $dagger$ superscript represents the hermitian
conjugate operation, namely complex conjugate plus transpose.

Thus the kind of operations allowed in duality computer are more general than those in
a quantum computer. It takes the general form $p_1U_1+p_2U_2$ for a two-path duality
computer where $U_1$ and $U_2$ are unitary operators and $p_1+p_2=1$.

If the input state of the duality computer is a mixed state, according to Gudder, the
divider operation is
\begin{eqnarray}
D_p\rho D_p^{\dagger}=\oplus p_i\rho,
\end{eqnarray}
namely the mixed state has been divided into $n$ parts. Then if the duality gate
operations, $\oplus U_i$ is applied to each path, the state becomes $\oplus p_i U_i\rho
U^{\dagger}_i$. Then the quantum wave combiner operation gives the following result
$\sum p_i U_i\rho U^{\dagger}_i$, namely the combiner operation in the density
formalism gives
\begin{eqnarray}
C_p(\rho_1\oplus\rho_2\cdots\oplus
\rho_n)C^{\dagger}_p=\sum_i \rho_i.
\end{eqnarray}
Because $\sum p_i U_i\rho U^{\dagger}_i$ describes a quantum operation, including both
unitary operation and measurements, the mixed state approach gave generally different
result from those obtained from the pure state approach, and a paradoxical situation
occured.

\section{Mathematics of Duality computer in the density matrix
formalism}.

We adopt Gudder's mathematical definition of divider operation \cite{r2}, however in
accordance with Eq. (\ref{divider})
\begin{eqnarray}
D_p\rho D_p^{\dagger}={1\over \|p\|^2}\oplus p_i\rho.
\end{eqnarray}
However for the combinor operation we define,
\begin{eqnarray}
C_p(\oplus p_i U_i\rho U_i^{\dagger}) C_p^{\dagger}=\|p\|^2(\sum_i p_i U_i)\rho (\sum_i
p_i U_i)^{\dagger}.\label{ecom}
\end{eqnarray}
It is easy to check that if the input state is a pure state, then the combined
operation of divider and combiner leaves the state invariant, \begin{eqnarray}
 C_pD_p & ( &|\psi\ket \bra \psi|)D_p^{\dagger} C_p^{\dagger}={1\over \|p\|^2}C_p(\oplus
 p_i|\psi\ket\bra\psi|)C_p^\dagger\nonumber\\
 &=&\sum_i(p_iI_H)|\psi\ket\bra\psi|(\sum_i p_iI_H)=|\psi\ket\bra\psi|.
\end{eqnarray}

The physical meaning is apparent if we interpret the density matrix in the proper
mixture sense\cite{r3}: we run the duality computer $N$ times and $N_i$ times the input
state is $|\phi_i\ket$, and then the density matrix for the input state  can be written
as $\rho=\sum_i q_i |\phi_i\ket\bra\phi_i|$ where $q_i=N_i/N$. Then in the $j$-th round
of duality computation, the resulting state is
$$\sum_i p_i U_i|\phi_j\ket.$$
In term of the density matrix formalism, it is \begin{eqnarray} (\sum_i
p_iU_i|\phi_j\ket\bra \phi_j| (\sum_k p_k U_k)^{\dagger}.\end{eqnarray}
After the $N$
rounds of duality computation, the averaged state after a duality computation is just
the average, namely
 \begin{eqnarray}
 &&\sum_j q_j(\sum_i p_iU_i|\phi_j\ket\bra \phi_j|
(\sum_k p_k U_k)^{\dagger}\nonumber\\
& =&(\sum_i p_iU_i)\rho (\sum_k p_k U_k)^{\dagger}.\end{eqnarray} Eq. (\ref{ecom})
indicates that the combinor operation can only combine sub-waves for the same quantum
system, i.e., all the sub-waves are from the same original wave through a quantum wave
divider. This is a natural requirement of the general principle of quantum interference
\cite{r2}.  Now we check the pure state formalism results of Gudder in the density
matrix formalism.

Lemma 2.1 is also valid in the density matrix formalism, i.e. the divider operation is
an isometry.

Lemma 2.2 is also true, i.e., the combiner operation $C_p$ is also an isometry.

Theorem 2.3 is  also valid here in the density matrix formalism, namely the identity
$I_H$ is an extreme point of ${ g}(H)$.

Corollary 2.4 is also true. Namely the extreme points of ${ g}(H)$ are precisely the
unitary operators of $H$.

Theorem 2.5 is also valid in the density matrix formalism, namely all bounded linear
operators on $H$ is in the positive cone generated by the convex set whose extreme
points are the unitary operators on $H$. The proof of this theorem in the density
matrix formalism can be made in the same way as in the pure state formalism of Gudder,
noting that the density $\rho$ can be written as a convex combination of pure state
density matrices.

Corollary 2.6 is also true in the density formalism.

\section{The measurement efficiency in the duality computer}

In Ref.\cite{r2}, we have pointed that there exists the problem of measurement
efficiency of a partial wave in the duality computer,  especially when the final result
is a near cancellation of the sub-waves. By measurement efficiency we mean the
probability to obtain a measured result if a perfect measuring apparatus is used.
Mathematically, these three cases can be described respectively by the following:

1) When measuring a partial wave, the probability distribution is the same as that when
measuring a full wave. In this case, this is equivalent to no renormalization of the
final quantum state, the final state after the duality computation in the pure state
formalism is
\begin{eqnarray}
 C_p
\oplus{1\over \|p\|} p_iU_i|\psi\ket=\sum_i(p_i U_i)|\psi\ket.
\end{eqnarray}
In general when an mixed input state is used, the final density matrix will be
\begin{eqnarray}
\sum_{i,j}(p_i U_i)\rho (p_j U_j)^{\dagger}.
\end{eqnarray}

2) In the second scenario, the measurement efficiency is 100\% but it may require  a
longer time to get a measured result. In the third scenario, the measurement efficiency
is 100\% and there is no time delay in getting a result. In both cases, they correspond
to a renormalization to the final state before the measurement provided that the norm
of density matrix is greater than a threshold depending on the measurement device,
\begin{eqnarray}
|\psi_{out}\ket=\left\{\begin{array}{ll}{\sum_i p_i U_i |\psi\ket \over |\bra
\psi|(\sum_i\bra p_i U_i)^{\dagger}(\sum_j p_i U_j)|\psi\ket|}, & |\sum_j p_i
U_j|\psi\ket|> \epsilon\\ 0, & else\end{array}\right.
\end{eqnarray}
where $\epsilon >0$ is a small number dependent on the measurement device. For an ideal
measurement device, $\epsilon=0$, namely as long as the sub-waves do not cancel
completely, the measurement device will detect it.

For mixed state, in terms of the density matrix formalism, this means that the
resulting density matrix just before the measurement is
\begin{eqnarray}
\rho_{out}={(\sum_i p_i U_i)\rho (\sum_j p_j U_j)^\dagger \over {\rm Tr}\left( (\sum_i
p_i U_i)\rho (\sum_j p_j U_j)^\dagger\right)},\label{efinal}
\end{eqnarray}
if ${\rm Tr}\left( (\sum_i p_i U_i)\rho (\sum_j p_j U_j)^\dagger\right)> \epsilon$. It
will be zero if ${\rm Tr}\left( (\sum_i p_i U_i)\rho (\sum_j p_j U_j)^\dagger\right)<
\epsilon$.

  Even in the first scenario, quantum Zeno effect may be used to increase the
measurement efficiency as one can repeatedly measure the same partial wave. Then this
may bring the final density matrix to the form in Eq. (\ref{efinal}).

For efficient calculation, the second and third scenario are preferred. However the
final choice is up to to Nature and it depends further experimental study. However,
even in the first scenario, the duality computer is at least as powerful as a quantum
computer because when there is only a single path, a duality computer is just a quantum
computer. With the help of quantum Zeno effect, the measurement efficiency can be
increased.

This work is supported by the National Fundamental Research Program Grant No.
001CB309308, China National Natural Science Foundation Grant Nos . 10325521, 60433050,
the Hang-Tian Science Fund, and the SRFDP program of Education Ministry of China.

\end{document}